\documentstyle[11pt,epsf]{article}
\begin{document}
%%%%%%%%%%%%%%%
\catcode`@=11
% Redefine caption to put text and formulas in smaller font
\long\def\@caption#1[#2]#3{\par\addcontentsline{\csname
  ext@#1\endcsname}{#1}{\protect\numberline{\csname
  the#1\endcsname}{\ignorespaces #2}}\begingroup
    \small
    \@parboxrestore
    \@makecaption{\csname fnum@#1\endcsname}{\ignorespaces #3}\par
  \endgroup}
\catcode`@=12
%%%%%%%%%%%%%%%%%%%%%%%%%%%%%%%%%%%%%%%%%%%%%%%%%%%%%%%%%%%%
\newcommand{\newc}{\newcommand}
\newc{\gsim}{\lower.7ex\hbox{$\;\stackrel{\textstyle>}{\sim}\;$}}
\newc{\lsim}{\lower.7ex\hbox{$\;\stackrel{\textstyle<}{\sim}\;$}}
%%%%%%%%%%%%%%%%%% Reference Defs %%%%%%%%%%%%%%%%%%
\def\NPB#1#2#3{Nucl. Phys. {\bf B#1} (19#2) #3}
\def\PLB#1#2#3{Phys. Lett. {\bf B#1} (19#2) #3}
\def\PLBold#1#2#3{Phys. Lett. {\bf#1B} (19#2) #3}
\def\PRD#1#2#3{Phys. Rev. {\bf D#1} (19#2) #3}
\def\PRL#1#2#3{Phys. Rev. Lett. {\bf#1} (19#2) #3}
\def\PRT#1#2#3{Phys. Rep. {\bf#1} (19#2) #3}
\def\ARAA#1#2#3{Ann. Rev. Astron. Astrophys. {\bf#1} (19#2) #3}
\def\ARNP#1#2#3{Ann. Rev. Nucl. Part. Sci. {\bf#1} (19#2) #3}
\def\MODA#1#2#3{Mod. Phys. Lett. {\bf A#1} (19#2) #3}
\def\ZPC#1#2#3{Zeit. f\"ur Physik {\bf C#1} (19#2) #3}
\def\APJ#1#2#3{Ap. J. {\bf#1} (19#2) #3}
%%%%%%%%%%%%%%%%%%%%%%%%%%%%%%%%%%%%%%%%%%%%%%%%%%%%%%%%%%%%%
\def\bdm{\begin{equation}}
\def\edm{\end{equation}}
\def\beq{\begin{eqnarray}}
\def\eeq{\end{eqnarray}}
\def\bea{\begin{eqnarray}}
\def\eea{\end{eqnarray}}
%%%%%%%%%%%%%%%%%%%%%%%%%%%%%%%%%%%%%%%%%%%%%%%%%%%%%%%%%%%%%%
\def\lqcd{\Lambda_{QCD}}
\def\lb{\Lambda_{b}}
\def\bwf{\Psi_{\lb}}
\def\lbp{\Lambda_{b} \rightarrow p\, l\, \bar \nu_{l}}
\def\vub{V_{ub}}
\def\bpi{B^0 \rightarrow \pi\,  l^+\, \nu_{l}}
\def\lom{{\lqcd \over m_{b}}}
\def\gf{\gamma_5}
\def\gmu{\gamma^{\mu}}
\def\gml{\gamma_{\mu}}
\def\pwf{\bar{\Psi}_{P}}
\def\dps{\lbrack d \xi_i \rbrack}
\def\pp{p^\prime}
\def\idc{\int_0^1{\dps}}
\def\ikm{\int {dk_2^- \over \sqrt2} \int {dk_3^- \over \sqrt2}}
\def\rme{\lbrack \bar u_P(\pp) \gml (1 - \gf) u_{\lb}(p) \rbrack}
\def\xar{(\xi_1,\xi_2,\xi_3)}
\def\kht{\hat k}
\def\kom{{\kht^2 \over \mu^2}}
\def\knm{{ {(2 \kht \cdot n)^2} \over {\mu^2} }}
\def\asm{\alpha_s(\mu)}
\def\psp{\bar u_P (\pp)}
\def\lbs{u_{\lb} (p)}
\def\dpp{\lbrack d^2 l_\perp \rbrack}
\def\dbt{\lbrack d^2 b \rbrack}
\def\tbl{\tilde b \lqcd}
%%%%%%%%%%%%%%%%%%%%%%%%%%%%%%%%%%%%%%%%%%%%%%%%%%%%%%%%

\begin{titlepage}
\begin{flushright}
{\large
UM-TH-95-11 \\
SPhT Saclay T95/049\\
hep-ph/9505378\\
\today\\
}
\end{flushright}
\vskip 2cm
\begin{center}
{\Large\bf Exclusive Semi-Leptonic Decays of b-Baryons into Protons}
\vskip 1cm
{\Large
Will Loinaz\footnote{E-mail: {\tt loinaz@walden.physics.lsa.umich.edu}}\\}
\vskip 2pt
{\large\it Randall Laboratory of Physics,\\
 University of Michigan, Ann Arbor, MI 48109--1120, USA}\\
{\Large
R. Akhoury\footnote{On sabbatical leave from the University of Michigan,
Ann Arbor}\\}
\vskip 2pt
{\large\it CEA Service de Physique Theorique, CE-Saclay\\
F-91191 Gif-sur-Yvette, CEDEX, France}\\
\end{center}

\vskip .5cm
\begin{abstract}
We present a QCD-based calculation of the exclusive semileptonic decay
$\lbp$.  Using the ideas of Heavy Quark Effective Theory, we discuss the
factorization of the amplitude.  Further, resummed Sudakov effects are put
in to ensure a consistent perturbative expansion.
\end{abstract}
\end{titlepage}
\setcounter{footnote}{0}
\setcounter{page}{2}
\setcounter{section}{0}
\setcounter{subsection}{0}
\setcounter{subsubsection}{0}
\setcounter{equation}{0}

%%%%%%%%%%%%%%%%%%%%%%%%%%%%%%%%%%%%%%%%%%%%%%%%%%%%%%%%%%%%%%%%%%%%%%%

The study of b-hadron decays has been the subject of considerable interest
in recent
years, as a source of information about CKM matrix elements, \cite{BS}
as a
laboratory for the application of QCD \cite{AC,Isgur,ASY,LiBmeson}
 and the development of theoretical
tools such as Heavy Quark
Effective Theory (HQET)\cite{Georgi}.  In particular, heavy-to-light decays are
interesting
because they give information on $\vub$, but they are especially difficult to
calculate because of the essential presence of strong interactions in the
hadronic bound state.  Recently, however, in \cite{ASY}
 a method has been formulated for analyzing
 the exclusive decay $\bpi$
in the region of large hadronic recoil and in the limit of a very heavy
b-quark.  The approach combines HQET, perturbative factorization theorems
for exclusive processes \cite{BrodskyLepage},
and exponentiation of Sudakov double-logarithms \cite{Mueller}.
  In
this paper we extend this technology to study the exclusive decay $\lbp$,
also in the limit of large hadronic recoil and large b quark mass.
In this limit, the method provides an asymptotic regime in which a systematic
expansion exists with corrections
${\cal O} ({\alpha}_s (c \lqcd m_b))$, where $c$ is a calculable constant.
For some
recent model-dependent calculations of $\Lambda_b$ decays, see
\cite{singleton}.
{}From the phenomenological side,to the extent that the very large $m_b$ limit
is realistic our
calculation is motivated by the possibility that it could provide another
method for extracting $\vub$ from data,
complementary to the B-meson studies. In addition, comparison of experiment
with the predictions of our calculation will afford added insight into the
applicability of perturbative QCD (PQCD) in processes with momentum transfers
in the few-GeV range.  Further, the analysis of this paper is useful for
contrasting
the
decays $\lbp$ and $\bpi$, which may be helpful in elucidating the surprisingly
large difference in the $\lb$ and $B^0$ lifetimes recently seen in experimental
data \cite{Moriond}.

We stress that in this paper we are calculating the purely perturbative
contribution
to the above-mentioned exclusive decay.  This would be the dominant one in the
limit of a very heavy b-quark.  However, for realistic values of $m_b$  there
are
could be sizable nonperturbative contributions in the form of higher-twist
effects
.  These will be discussed in a future work \cite{future}.  However, as we
shall see
below, numerical indications are that a nodified perturbation expansion is
self-consistent even for realistic values of $m_b$.

The physical picture of this decay is similar to that for the meson, described
in \cite{ASY}.  Sitting inside the $\lb$, the b-quark decays into a $W^-$
and a fast-moving, nearly on-shell u-quark.  The u-quark propagates through
the remaining hadronic medium, picking up a light $ud$ pair over a distance
$y$.
Since we are considering an exclusive decay, no gluon radiation escapes.  If
$y$
were large enough  (${\cal O}({1 \over \lqcd} )$) we would expect
considerable gluon
radiation.  Thus, for exclusive processes we expect the outgoing u-quark to
propagate only small distances ($<< {\cal O}({1 \over \lqcd})$ )
before acquiring a $ud$ pair
to form the color singlet which will hadronize to form the proton.  Then the
hard gluons
which kick these spectator quarks will typically be off-shell by
${\cal O}(\lqcd m_{b})$,
large enough that we might expect perturbative factorization theorems to
apply.

We will state the arguments and results of our approach and sketch the
calculations.  The
calculational details will be published elsewhere \cite{future}.
Our procedure is to judiciously combine elements of \cite{ASY} and
\cite{LiSterman,Liproton}.
  We first identify the sources of
long-distance behaviour which must be dealt with in order to successfully
apply
perturbation theory.  Soft divergences will arise from interactions of soft
gluons with the heavy quark, and soft and collinear divergences will arise
from the interactions of virtual partons travelling in the direction of the
outgoing proton.  As in \cite{ASY}, we separate out the soft divergences due to
the
heavy quark
using an eikonal treatment of the b-gluon interactions.  All the other
divergences
we regulate using the Sudakov resummation procedure developed in
\cite{BottsSterman}
and applied to exclusive decays in \cite{LiSterman,Liproton}.
Of course these do not include the
collinear divergences that are collected into the universal proton
wavefunction.
This modification of the perturbative expression takes into account the
transverse
momenta of the partons inside the proton.  The resummation of the most
important
contribution behaves like
$exp[-const \times \log Q ( \log ({{\log Q} \over {\log b}}) )]$
in the leading logarithmic approximation where $b$ is the conjugate Fourier
transform
variable to $l_\perp$.  In the above $Q$ is a typical momentum transfer which
is of
the order of $m_b$ in our case.  This Sudakov suppression selects the
components of
the proton wavefunction with small transverse separation, thereby ensuring that
the exchanged hard gluon is always offshell and consequently
perturbation theory for the
hard scattering amplitude remains stable and reliable.
Contributions from additional non-valence Fock states will be
suppressed by ${\cal O}(\lom)$ from standard parton model considerations
for hard exclusive processes \cite{BrodskyLepage}.

We conclude by calculating the form factor and
differential decay rate, combining our form factor with model $\Lambda_{b}$
and proton wavefunctions.

We now discuss the factorization of the hadronic form factor.
The interaction Hamiltonian relevant to the decay $\lbp$ is

\bdm  H_{int}={G_F \over \sqrt2} V_{ub} {( \bar u \gml (1 - \gf )
  b )} {( \bar l \gmu ( 1 - \gf )  \nu_l )} \edm

The amplitude is
\bdm
{\cal M} ={G_F \over \sqrt2} V_{ub} \underbrace{\langle P(\pp) | \bar u \gml
(1 - \gf ) b | \lb (p) \rangle}_{M_{\mu}} {( \bar l  \gmu ( 1 - \gf )
\nu_l )}
\edm

We will be concerned with calculating the hadronic matrix element $M_{\mu}$
in the limit of $x \sim 1$, where $x$ is defined by
\bdm
x={{2 p \cdot \pp} \over {m_b}^2}
\edm
where $p$ is the momentum of the $\lb$ and $\pp$ is that of the outgoing
proton,
which we take to be in the $(+)$ direction.
In \cite{ASY} it was shown that for $\bpi$ the hadronic matrix element could be
sensibly written as a
convolution of a hard-scattering amplitude with initial- and final- state
 valence
hadronic wavefunctions.  For the present case the analogous treatment
is valid. As we discuss below, we can write (suppressing color and spinor
indices):

\bdm
M_{\mu}(p,\pp)= \int{d^4 k_{2}\over (2 \pi)^4} {d^4 k_{3}\over
(2 \pi)^4} {\cal A}_{\mu}(p,p^\prime) \bwf (p,k_{2},k_{3})
\edm
where $k_2$ and $k_3$ are light quark initial momenta, $\bwf$ is the
$\lb$ valence wavefunction, and $\cal A_{\mu}$ is itself a convolution of the
hard-scattering amplitude, $H_{\mu}$, and the proton valence wavefunction,
$\pwf$.
Written explicitly,
\bdm
M_{\mu}(p,\pp)=
\int_0^1 \dps \int \dpp \int{d^4 k_{2}\over (2 \pi)^4} {d^4 k_{3}\over
(2 \pi)^4} \pwf (\xi_i,\pp,l_{\perp i}) H_{\mu} (\xi_i,l_{\perp
i},\pp,p,k_2,k_3)
\bwf (p,k_2,k_3)
\edm

\bdm
\dps = {\delta \left (1 - \sum_{i=1}^{3} \xi_i \right) d \xi_1 d \xi_2 d \xi_3}
\edm
\bdm
\dpp = {\delta}^2 \left( \sum_{i=1}^3 l_{\perp i} \right) d^2 l_{\perp 1} d^2
l_{\perp 2}
 d^2 l_{\perp 3}
\edm
where the $\xi_{i}$ are the longitudinal momentum fractions of the proton's
valence quarks and $l_{\perp i}$ are the corresponding perp components.

We now briefly outline the steps leading to eqn.[4].
As usual in the treatment of exclusive processes, to yield a hard-scattering
amplitude computable in PQCD the infrared divergences must be factored from
the hard-scattering amplitude into the wavefunctions.  The factorization
procedure is a straightforward extension of the corresponding analysis in
\cite{ASY}.  We review the salient points of the factorization of soft
divergences
here.  The remaining collinear divergences will be controlled by the
resummation procedure discussed later.

\begin{description}
\item[Step 1] Decoupling assumption:  integrate out the heavy b-quark
loop effects, allowing the matrix element to be written as
\bdm
M_{\mu}= {\langle P(\pp)| J_{\mu} (0) | \lb \rangle}_{{\cal L}_{QCD}}
={\langle P(\pp)| T(J_{\mu} (0) V(A^0) ) | \lb \rangle}_{{\cal L} (q)}
\edm
Here $V$ represents a single b-quark line connected to the weak vertex.  $V$ is
 composed of a sequence of b-gluon vertices and b-propagators and as such
contains the full effect of the terms of the Lagrangian
${\cal L}_0 (b) - i g \bar b \gamma \cdot A b $, up to loop corrections.  It is
$V$
which now contains the soft divergences.  Note that the matrix element is now
calculated using the light-quark Lagrangian density
${\cal L} (q)$ only.
\item[Step 2] Go to the $\lb$ rest-frame and introduce the eikonal phase.
\bdm
U(A^{0})= {\cal P} exp{\lbrack -i g \int_{-\infty}^0{ \!\! n \cdot A(\lambda
n_{\mu})
d\lambda}}
\rbrack
\edm
$n$ is a unit vector in the direction of the heavy-quark velocity
(here $n_{\mu}=\delta_{0 \mu}$).
This contains precisely the soft divergences of $V(A^0)$ that we wish to
remove.
The associated Feynman rule is shown explicitly in \cite{ASY}.
\item[Step 3] Define a subracted b-quark line, $V^{fin}=VU^{-1}$,
from which the soft divergences have been removed.  To lowest order this
 is the difference between the full b-quark propagator and the eikonal
propagator.
\item[Step 4] Express the hadronic matrix element in terms of this subtracted
propagator, inserting a $UU^{-1}$ and a sum over a complete set of states to
obtain:
\bdm
M_{\mu}= \sum_{n=3}^{\infty} {\langle P(\pp)| T(J_{\mu} (0) V^{fin} (A^0)
\otimes_n U(A^0 )) | \lb \rangle}_{{\cal L} (q)}
\edm
Here the $\sum_{n=3}^{\infty} \ldots \otimes_n \ldots$ denotes a summation
over intermediate states with $n-1$ light partons, integration over their
momenta, and summation over other relevant indices.
\item[Step 5] Neglect the higher Fock states.  What remains after a
reorganization of the perturbative sum is:
\bdm
M_{\mu}= \int{d^4 k_{2}\over (2 \pi)^4} {d^4 k_{3}\over
(2 \pi)^4} \sum_{n=3}^{\infty} \lbrack {\langle P(\pp) | T(J_{\mu} (0)
V^{fin} (A^0))
| bud \rangle} {\langle bud | U(A^0 )) | \lb \rangle} \rbrack_{{\cal L} (q)}
\edm
where we identify the second factor in the integrand as the
$\lb$ wavefunction and the first factor as $\cal A_{\mu}$ from eqn (3).
The passage to eqn (4) is justified if the process the proton emerges from a
short-distance region
and the procedure is by now fairly standard (see \cite{Liproton}).
\end{description}

We next discuss the implementation of the Sudakov suppression a la' Li and
Sterman.
For this purpose, following \cite{LiSterman,Liproton} we work in an
axial gauge.  The first step
is to transform eqn (4) into '$b$' space, which is Fourier-conjugate to
$l_{\perp}$:

\bdm
M_{\mu}(p,\pp)=
\int_0^1 \dps \int \dbt \int{d^4 k_{2}\over (2 \pi)^4} {d^4 k_{3}\over
(2 \pi)^4} {\cal P}_{p} (\xi_i,\pp,b_i,\mu)
H_{\mu} (\xi_i,\pp,p,k_2,k_3,b_i,\mu)
\bwf (p,k_2,k_3)
\edm
where
\bdm
\lbrack d^2 b \rbrack = { {d^2 b_2} \over (2 \pi)^2 }
{ {d^2 b_3} \over (2 \pi)^2 }
\edm
$\mu$ is the factorization scale, and ${\cal P}_{p} (\xi_i,\pp,b_i,\mu)$
are the transformed proton wavefunctions.  The $\cal P$s include all the large
logarithmic and double logarithmic radiative corrections at large $b$.  These
have
been resummed and exponentiated \cite{BottsSterman} and result in a suppression
of
the distribution amplitude at large $b$.  Explicitly, we write
\bdm
{\cal P}_{p} (\xi_i,\pp,b_i,\mu)=
exp \left[- \sum_{l=1}^3 \left( s(\xi_{l}, \tilde{b_l}, m_b) + \int_{1 \over
{\tilde{b}_l}}
^\mu { d\bar \mu \over {\bar \mu} } \gamma_q (g^2(\bar \mu)) \right) \right]
\pwf(\xi_i,\pp, w)
+ {\cal O}(\alpha_s^2(w))
\edm
where $w={\rm min}_i({1 \over {\tilde{b}_i} })$.  The $s(\xi_l,\tilde{b}_l,
m_b)$ are
defined in \cite{BottsSterman,Liproton}
and $\gamma_q = - {{\alpha_s} \over \pi}$ is the
axial gauge quark anomalous dimension.  The $\tilde{b_l}$ are the infrared
cutoff
parameters and we follow the "MAX" prescription of \cite{bolz}, i.e.
\bdm
\tilde{b}_l={\rm max}[b_1,b_2,b_3]
\edm
where $b_1=|b_2 - b_3|$.
Including the $\mu$ dependence of the hard-
scattering amplitude we can write our final result in the form
\bea
M_{\mu}(p,\pp)&=& {1 \over {(2 \pi)^3}}
\int_0^1 \dps \int_0^{\infty} b_2 db_2 \int_0^{\infty} b_3 db_3 \int_0^{2 \pi}
d\theta
 \int{d^4 k_{2}\over (2 \pi)^4} {d^4 k_{3}\over
(2 \pi)^4} \nonumber  \\
& & {\pwf} (\xi_i,\pp,w)
 H_{\mu} (\xi_i,b_i,\theta,\pp,p,k_2,k_3,t_{\alpha_1},
t_{\alpha_2} ) \bwf (p,k_2,k_3) \nonumber  \\
& & \times exp\left[ - S(\xi_i,\tilde{b_i},m_b,
\tilde{t}_{\alpha_1},\tilde{t}_{\alpha_2}) \right]
\eea
where $\theta$ is the angle between $b_2$ and $b_3$.
  In the above, $t_{\alpha_1}, t_{\alpha_2}$
are the appropriate
scales at which the running couplings in the hard scattering amplitude are
evaluated
and are related to the largest mass scales appearing in $H_\mu$.
The detailed forms are given below.
The $S$ are generically of the form
\bdm
S=\sum_{l=1}^3 \left[ s(\xi_l,\tilde{b}_l,m_b) - \int_{1 \over
\tilde{b}_l}^{\tilde{t}_l}
 { d\bar \mu \over {\bar \mu} } \gamma_q (g^2(\bar \mu)) \right]
\edm
where the $\tilde{t}_l$ depend on the hard scattering diagram (see below).

We now express the wavefunctions in explicit forms more useful
for calculations.  In the above we have neglected any intrinsic $l_\perp$
dependence
in the wavefunctions.  A more complete treatment taking this into account will
be given
elsewhere \cite{future}.
We work in the rest frame of the $\lb$ with the proton
moving off in the $(+)$ direction.  Following the notation of other authors
\cite{CZ,KS} we write the final-state proton wavefunction:

\bea
\pwf (\xi_i)_{\alpha \beta \gamma , l m n}& = &{1 \over 6}
\epsilon_{lmn} {1 \over 4} f_P
[ ( C \gamma \cdot \pp)_{\alpha \beta}
(\psp \gf)_\gamma V(\xi_1 , \xi_2 , \xi_3) \nonumber \\
& & \hspace{.75in}+ ( C \gamma \cdot \pp \gf )_{\alpha \beta}
\psp_\gamma A(\xi_1 , \xi_2 , \xi_3) \nonumber \\
& & \hspace{.75in}+ (i C \sigma_{\delta \omega} \pp_\omega ) (\psp
\gamma_\delta \gf )_\gamma T(\xi_1 , \xi_2 , \xi_3)  ]   \\
&=& \int {\prod_{i=1}^3 \left( {d(z_i \cdot \pp) \over (2 \pi)} e^{i k_i \cdot
z_i} \right)
\langle P(\pp)| {\bar u}_\alpha ^l (z_1) {\bar u}_\beta ^m (z_2)
{\bar d}_{\gamma} ^n (z_3) | 0 \rangle  }
\eea
where
\bea
 2 T(\xi_1,\xi_2,\xi_3)=\phi_P (\xi_1,\xi_3,\xi_2) +
\phi_P (\xi_2,\xi_3,\xi_1)     \\
 2 V(\xi_1,\xi_2,\xi_3)= \phi_P (\xi_1,\xi_2,\xi_3) +
\phi_P (\xi_2,\xi_1,\xi_3)     \\
 2 A(\xi_1,\xi_2,\xi_3)= \phi_P (\xi_2,\xi_1,\xi_3) -
\phi_P (\xi_1,\xi_2,\xi_3)
\eea
and $\xi_1 + \xi_2 + \xi_3 = 1$. $C$ is the charge conjugation operator and
$\sigma_{\mu \nu}=
{i \over 2} \lbrack \gamma_{\mu},\gamma_{\nu} \rbrack$.
$f_P$ is related to the wavefunction at the origin.  It is chosen such that
the distribution amplitude has normalization
$\int_0^1 \dps { \, \phi_P (\xi_i ,\mu^2)}=1$.

We may define the $\lb$ wavefunction similarly:
\bea
\bwf (k_2,k_3)_{\alpha \beta \gamma, l m n}=
{1 \over 6} \epsilon_{lmn} \left[
{\left[ (\gamma \cdot p + m_b) \gf C \right]}_{\beta
\gamma} u(p)_{\alpha} \psi (p,k_2,k_3)  \right]              \\
=  \int {d^4 z_2}  e^{i k_2 \cdot z_2}
\int {d^4 z_3}  e^{i k_3 \cdot z_3}
\langle 0 | b_{\alpha}^l (0) U(A^0) u_{\beta}^m (z_2) d_{\gamma}^n (z_3)
| \lb(p) \rangle
\eea

Note that our choice of a scalar momentum wavefunction $\psi(p,k_2,k_3)$
(rather than
a more complicated tensor structure) corresponds to an assumption of
decoupling of the spin and orbital degrees of freedom
in the light-quark system, as discussed in \cite{HK}.

With our choice of frame we find that to leading order in $m_b$ the only
nontrivial dependence of the hard-scattering amplitude on $k_2$ and $k_3$
is through their $(-)$ components.  Thus it is convenient to define a
distribution amplitude $\tilde{\phi} (k_2^-,k_3^-)$ which replaces
$\psi(p,k_2^-,k_3^-)$ and in which the $(+)$ and $(\perp)$
components have been integrated out:
\bea
\tilde{\phi} (k_2^-,k_3^-) =
\int {{dk_2^+ d^2 k_{2 \perp}}\over {(2 \pi)^4 \sqrt2}} \int {{dk_3^+ d^2 k_{3
\perp}}
 \over {(2 \pi)^4 \sqrt2}}
 \psi(p,k_2,k_3)
\eea
where $k_i^{\pm}=k_i^0 \pm k_i^3$.
It is convenient to change to the dimensionless variables
 $\xi, \eta$
defined by $\xi={k^-_2 \over {k^-_2 + k^-_3}}$
and $\eta={{k^-_2 + k^-_3} \over {m_b}}$.
In terms of these variables we normalize the
distribution amplitude $\phi (\xi,\eta)$ by
\bdm
\int_0^1 d\xi \int_0^1 \eta d\eta \phi (\xi,\eta)=1
\edm
and define $f_{\lb}$ analogous to $f_p$:
\bdm
\bwf (\xi,\eta)_{\alpha \beta \gamma, l m n}=
{1 \over 6} \epsilon_{lmn} \left[
{1 \over 4} f_{\lb} {\left[ (\gamma \cdot p + m_b) \gf C \right]}_{\beta
\gamma} u(p)_{\alpha} \phi (\xi,\eta)  \right]
\edm

Thus, $f_{\lb}$ is related to the $\lb$ wavefunction at the origin
to leading order in $\alpha_s(m_b^2)$.  The higher order effects can be
discussed;
see \cite{AR} for the meson case.
Using the above definitions and evaluating the traces, the Dirac structure of
the
leading terms
will be of form $\rme$. The matrix element may then be written (with color
indices and color wavefunctions suppressed):
\bea
M_{\mu}&=& {1 \over 2} {1 \over (2 \pi)^3} {({1 \over 4})}^2 f_{\lb} f_P \rme
\times \nonumber\\
& &
\idc \int_0^1 \!\!{\eta d\eta} \int_0^1 \!\!d\xi \int_0^\infty \!\!b_2 db_2
\int_0^\infty
 \!\!b_3 db_3
\int_0^{2 \pi} \!\!d\theta \bar{H}(\xi_i,
\eta,\xi,b_2,b_3,\theta,\pp,p,t_{\alpha_1},
t_{\alpha_2}) \times \nonumber \\
& & e^{-S} \phi(\xi,\eta)        \\
&=&\rme F(x).
\eea
Explicit expressions for $\bar{H} e^{-S}$ for the different diagrams are given
below.
Note that the proton wavefunctions are included in the definition of $\bar{H}$.

We now discuss the calculation of the hard-scattering amplitude, $H_\mu$.
Unlike the $\bpi$ case, for $\lbp$ only a small subset of the full set
of 16 tree-level diagrams contribute to leading order in $m_b$.  In a spacelike
axial
gauge with the gauge-fixing vector having no perp component,
the leading diagrams are those shown in figure [1].  After
contracting all spinor indices with the wavefunction
spin-projection operators and all color indices to give initial- and final-
state color singlets, and after manipulating integration variables,
their contributions to $\bar{H} e^{-S}$ are:
\begin{eqnarray}
\bar{H}^{(a1+a2)} & = &  {-16 (2 \pi)^4 C_B^2 \alpha_s(t_{22}^2)
\alpha_s(t_{33}^2)}
{\xi \over {\eta ( \xi \! - \! \xi_3)}}
{\left( V(\xi_1 \xi_3 \xi_2) + A(\xi_1 \xi_3 \xi_2) + 2 T(\xi_1 \xi_2 \xi_3)
\right)}
\times
\nonumber   \\
&  & K_0 ( \sqrt{\xi_2 x m_b^2 \xi \eta} b_2 )
 \! \left[ K_0(\sqrt{\xi_3 x m_b^2 (1 \!- \!\xi) \eta} b_3) \! - \!
K_0(\sqrt{(1 \! - \! \xi_3) x m_b^2 \xi \eta} b_3) \right]   \\
S^{(a)} & = & \sum_{i=1}^3 {s(\xi_i,{\tilde b)}} +
{1 \over {2 \beta_1}}
\left[ \log \left[ {{\log \tilde{t}_2}\over{- \hat b}}
\right] +
         \log \left[ {{\log \tilde{t}_3}\over{- \hat b}}
\right]
   +    \log \left[ {{\log \tilde{t}_6}\over{- \hat b}}
\right] \right]   \\
\bar{H}^{(b1+b2)}& = & {{64 (2 \pi)^4 C_B^2
\alpha_s(t_{12}^2) \alpha_s(t_{31}^2)} \over {(\xi_2 + \xi_3)}}
{1 \over {\eta (1 - \xi)}} {\left( V(\xi_1 \xi_2 \xi_3) + A(\xi_1 \xi_2 \xi_3)
+ 2 T(\xi_1 \xi_2 \xi_3) \right)} \nonumber \\
& & K_0(\sqrt{(\xi_2 + \xi_3) x m_b^2 \eta} b_3)
\lbrack K_0(\sqrt{\xi_2 x m_b^2 \xi \eta} b_1)-K_0(\sqrt{\xi_2 x m_b^2 \eta}
b_1) \rbrack  \\
S^{(b)} & = & \sum_{i=1}^3 {s(\xi_i,{\tilde b})} +
{1 \over {2 \beta_1}}
\left[ \log \left[ {{\log \tilde{t}_1}\over{- \hat b}} \right] +
 \log \left[ {{\log \tilde{t}_3}\over{- \hat b}}  \right]
 + \log \left[ {{\log \tilde{t}_4}\over{- \hat b}} \right] \right]  \\
\end{eqnarray}
where
\bea
t_{1i}&=&{\rm max} \lbrack \sqrt{(\xi_2 + \xi_3) x m_b^2 \eta} , {1 \over {b_i}
}
\rbrack \\
t_{2i}&=&{\rm max} \lbrack \sqrt{\xi_2 x m_b^2 \eta \xi} , {1 \over {b_i}}
\rbrack \\
t_{3i}&=&{\rm max} \lbrack \sqrt{\xi_3 x m_b^2 \eta (1 - \xi)}, {1 \over {b_i}}
\rbrack
\eea
and
\bdm
\tilde{t}_{1}= {1 \over {\lqcd}} {\rm max} \lbrack \sqrt{(\xi_2 + \xi_3) x
m_b^2 \eta} ,
{1 \over {\tilde b}} \rbrack
\edm
The remaining $\tilde{t}_{i}$ are defined similarly, as the maximum of
$\tilde{b} ^{-1}$ and the argument of one of the Bessel functions in
the associated $\bar{H}$.
Here $\beta_1={{33- 2 n_f} \over 12}$, ${\hat b}=\log ({\tilde b} \lqcd)$,
and $K_0$ is the modified Bessel function.

The remaining diagrams are either identically zero due to color algebra, as in
the case of the 3-gluon-vertex diagrams, or are suppressed by ${\cal O}(\lom)$
due to Dirac algebra. Note in particular that all diagrams involving a
subtracted heavy quark propagator (henceforth called slashed diagrams)
are suppressed by at least ${\cal O}(\lom)$ with
respect
to the leading diagrams.  It's instructive to contrast this with the meson
case to see why.  Consider first
 the meson the two tree-level diagrams are shown in
figure [2].
The slashed diagram naively appears suppressed with respect to the
other because
the heavy quark propagator is
${{ \gamma \cdot \pp } \over {{\pp}^+ p^-}} \sim {\cal O}({1 \over m_b})$
whereas the light-quark
propagator of the other diagram is ${{\gamma \cdot (\pp - k)} \over
{- {\pp}^+ k^-}}$ which is naively ${\cal O}({1 \over \lqcd})$.  The $\pp$ in
the
light-quark
numerator is eliminated by the final-state spinors, however,
leaving ${{ \gamma
\cdot  k} \over {{\pp}^+ k^-}} \sim {\cal O}({1 \over m_{b}})$ --  the same
order as
the slashed propagator.  For the baryon, consider the diagrams of
figure [3]. Analysis of these diagrams is simplified by the fact that
their leading terms come from the $g_{\mu \nu}$ pieces of the gluon
propagators in this choice of axial gauge.
Compare the leftmost quark propagators on the top line.
Once again the slashed
propagator appears to be suppressed.  This time the leading piece $\pp$ of the
leftmost light-quark propagator in the second diagram
is not eliminated, since the adjacent
light-quark
propagator is offshell and shields it from the final-state spinors.
Thus our naive counting of powers of $m_b$ remains correct and the slashed
diagram is suppressed with respect to the unslashed diagram, in contrast to
the meson case.
Combining eqns. (22),(23) with (24)-(28), we may obtain the expression for the
form factor $F(x)$.  This is the only dynamical difference that we find between
the meson and baryon formfactor calculations.

Next we proceed to the calculation of the differential decay rate.
Neglecting lepton and proton masses, the differential rate is:
\bdm
\left| {{d \Gamma} \over {d x}} \right| = {{G^2_F m^5_b} \over {96 \pi^3}} x^2
(3-2 x)
|F (x)|^2 |V_{ub}|^2
\edm

To proceed further we need models for the proton and $\lb$ wavefunctions.
Several models for the proton wavefunction have been proposed in the
literature.  We adopt the model of Chernyak and Zhitnitskii \cite{CZ}
 (neglecting
$Q^2$ evolution):

\bdm
\phi_P^{CZ}(\xi_i,\mu = 1 {\rm GeV}) = 120 \xi_1 \xi_2 \xi_3
\lbrack 11.35 (\xi_1^2 + \xi_2^2) + 8.82 \xi_3^2 - 1.68 \xi_3
-2.94 - 6.72 (\xi_2^2 - \xi_1^2) \rbrack
\edm

There is little data available on heavy baryon wavefunctions, but some
plausible models for $\Lambda$-baryons with unequal mass constituents have
been proposed from which we choose the following \cite{Schlumpf}:

\bdm
 \phi(\xi,\eta) = N \eta^2 (1- \eta) \xi (1- \xi) exp \left[
- {{m_b^2} \over {2 \beta^2 (1 - \eta)}} - {{m_l^2} \over {2 \beta^2 \eta
\xi (1-\xi)}  }  \right]
\edm
N is a normalization constant chosen to fulfill eqn.(26) and $m_l$ is the
light constituent quark mass.  In the above,
$\beta$ is a parameter related to the string tension and is to be fitted to
data.  We present results with two typical values for this parameter.

{}From the form of the
distribution amplitudes give above and expressions (28),(30), and (32), we
see that there are no remaining singularities from any endpoint regions.

Next the differential decay rate may be evaluated numerically for different
 values of $x$. To show the variation of ${d\Gamma} \over {dx}$ with $x$
, we write it in the form
\bdm
\left| {{d\Gamma} \over {dx}} \right|= |V_{ub}|^2 f_{\lb}^2 f_p^2 R_p(x)
\edm
$R_p(x)$ is plotted against $x$ for $x$ near 1 in figure [4]
for two different values of $\beta$, $\beta = 0.5$ GeV (solid line)
and $\beta= 1$ GeV
(dashed line).
In Table I we list the partial decay rates for $x$
integrated from $x_1$ to $x_2$.  To get
an order of magnitude estimate we use
$V_{ub} \sim 0.003$, $f_p=5.3 \times 10^{-3} {\rm GeV}^{2}$ \cite{CZ},
 and $f_{\lb} \sim f_p$.
The corresponding numerical
values for the partial rates are given in the last column of Table I.
The total width of $\lb$ is approximately $1.6 \times 10^{-12}$ GeV
\cite{PDG}, so that the perturbative partial widths calculated here are
of order $10^{-8}$ of the total $\lb$ decay width.
This is somewhat smaller than the branching ratios for the analogous
meson process $\bpi$ calculated in \cite{ASY}.  We stress, however, that
there is considerable uncertainty in the form and parameters of both the
$\lb$ and proton wavefunctions (see e.g. \cite{Gari}) and in the
value of $f_{\lb}$, which could have considerable impact on
our numerical results.  Thus our numerical values for the decay
rates should only be considered to be rough indicators.

The branching ratio for
$\lb \rightarrow \Lambda_c l \bar \nu_l$ is estimated to be approximately
$10^{-2}$ from model calculations \cite{singleton}.  Accounting for the
difference in the CKM elements and scaling down from the heavy-heavy
decay, we can get an order of magnitude estimate for the branching ratio
of $\lbp$ of $10^{-8}$, which is consistent with our result
given the uncertainties.

We next turn to a discussion of the numerical analysis.
We have investigated numerically the contributions of different regions in
$b$-space
to determine whether the perturbative contribution is dominant.  Following
the analysis of \cite{Liproton} we look at the contributions to the form factor
$F(x)$ from different parts of the integration region.
This is done by checking to see
if the dominant contribution comes from the small-$b$ regions of $b$-space.  We
check this
numerically by cutting off the each of the $b$-integrals at some maximum $b_c$,
i.e $b_1,b_2,b_3 < b_c$.  Our results are shown in figure [5] for $\beta=1 {\rm
GeV}$.  We
observe that for $\lqcd b_c > 0.8$ the value of $F(x)$ is essentially constant,
indicating that most of the contribution to the integral comes from the
region $b_1,b_2,b_3 < {{0.8} \over {\lqcd}}$.  The curves with the
largest values of $x$ appear to flatten more quickly than the smaller-$x$
curves at
large $b_c$, indicating that for more energetic outgoing protons the
perturbative
regime is more dominant than that of less energetic ones, as expected.

Following \cite{Liproton}, we may define $b_{1/2}$ to be the value of $b_c$
at which $50\%$ of the total value of $F(x)$ has been accumulated.  A
loose criterion proposed for the consistency of perturbation theory that
$\alpha_s^2 ({1 \over {b_{1/2}^2}}) < 1$.  Our result satisfies this criterion:
for $x=1, \beta=0.5$ GeV, $b_{1/2}$ is 0.40 and $\alpha_s=0.83$.
Other $x$ and
$\beta$ tested give larger values of $b_{1/2}$, but even the case
$\beta=1.0$ GeV, $x=0.75$ gave $\alpha_s=0.91$.  This indicates that most of
the
value of $F(x)$ is accumulated in the short-distance
region of the integration space in which
we may expect the perturbative expansion to be self-consistent.

In this paper we have calculated the perturbative contribution to the decay
$\lbp$ to leading order in ${1 \over {m_b}}$.
This is the dominant contribution as $m_b \rightarrow \infty$,
and we have investigated the self-consistency of the perturbative approach
for realistic values of $m_b$.  With our choice of model
wavefunctions for the heavy baryon and proton
we find that the perturbative contribution of
$\lbp$ to the total width of $\lb$ is of order $10^{-8}$ of the
total width.  Our confidence in the self-consistency of the
the result is supported by the fact that the largest contribution to
the form factor comes from the short-distance regions of the
integration space, as discussed above.  We stress again that there is
considerable uncertainty in the numerical results due to uncertainty
about the wavefunctions.  However, for at least one plausible set
 of model wavefunctions we obtain values for the decay rates comparable
to those calculated for $\bpi$ using a perturbative expansion
made consistent by the introduction resummed Sudakov effects.
The issue of the contributions from higher orders in ${1 \over {m_b}}$
(higher twist) is beyond the scope of this paper.

We are very grateful to George Sterman for many helpful conversations and
useful suggestions.  We would also like to thank Graham Kribs for help with
the computer programming.  One of us (R.A.) would like to thank the Service
Physique Theorique, Saclay, for hospitality and support.  This work was
supported in part by the U.S. Department of Energy.
\vspace{.6in.}

\vfill\eject

%%%%%%%%%%%%%%%%%%%%%%%%%%%%%%%%%%%%%%%%%%%%%%%%%%%%%%%%%%%%%%%%%%%%%%%
\newcounter{bean}
\setcounter{bean}{0}
\begin{center}
Figure Captions
\end{center}
\vspace{0.5in}

\begin{list}%
{Figure \arabic{bean}}{\usecounter{bean}
\setlength{\rightmargin}{\leftmargin}}
\item  Leading Tree-Level Diagrams in Spacelike Axial Gauge
\item  Tree-level Diagrams for $B^0 \rightarrow \pi l^+ \nu_l$
\item  A Suppressed Diagram and a Leading Diagram
\item  $R_p(x)$ vs. $x$ for $\beta=1$ GeV (dashed line) and
	$\beta=0.5$ GeV (solid line). $R(x)$ is in units of
	${\rm GeV}^{-7}$
\item	$40 F(x)$ vs. $b_{cutoff}$ for $\beta=1$ GeV.  The
	curves are $x=0.75,0.8,0.85,0.9,0.95,1.0$, intersecting
	the $b=1$ line from top to bottom respectively.  $b_{cutoff}$
	is in units of ${1 \over {\Lambda_{QCD}}}$
\end{list}
\vfill\eject
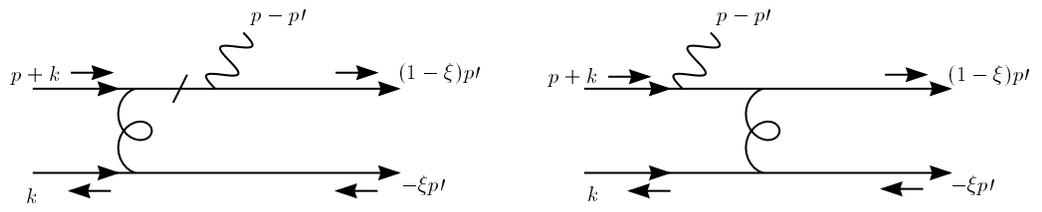
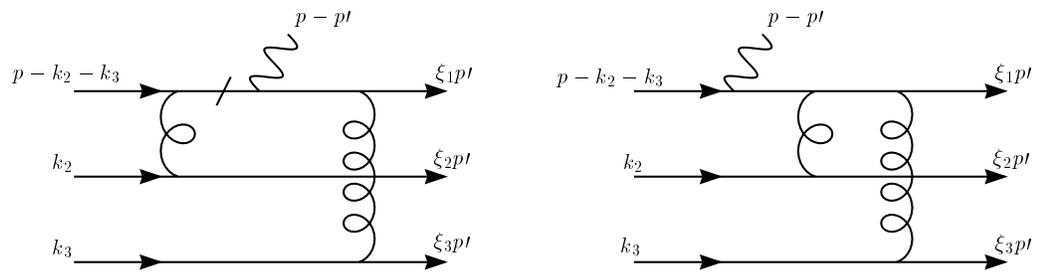
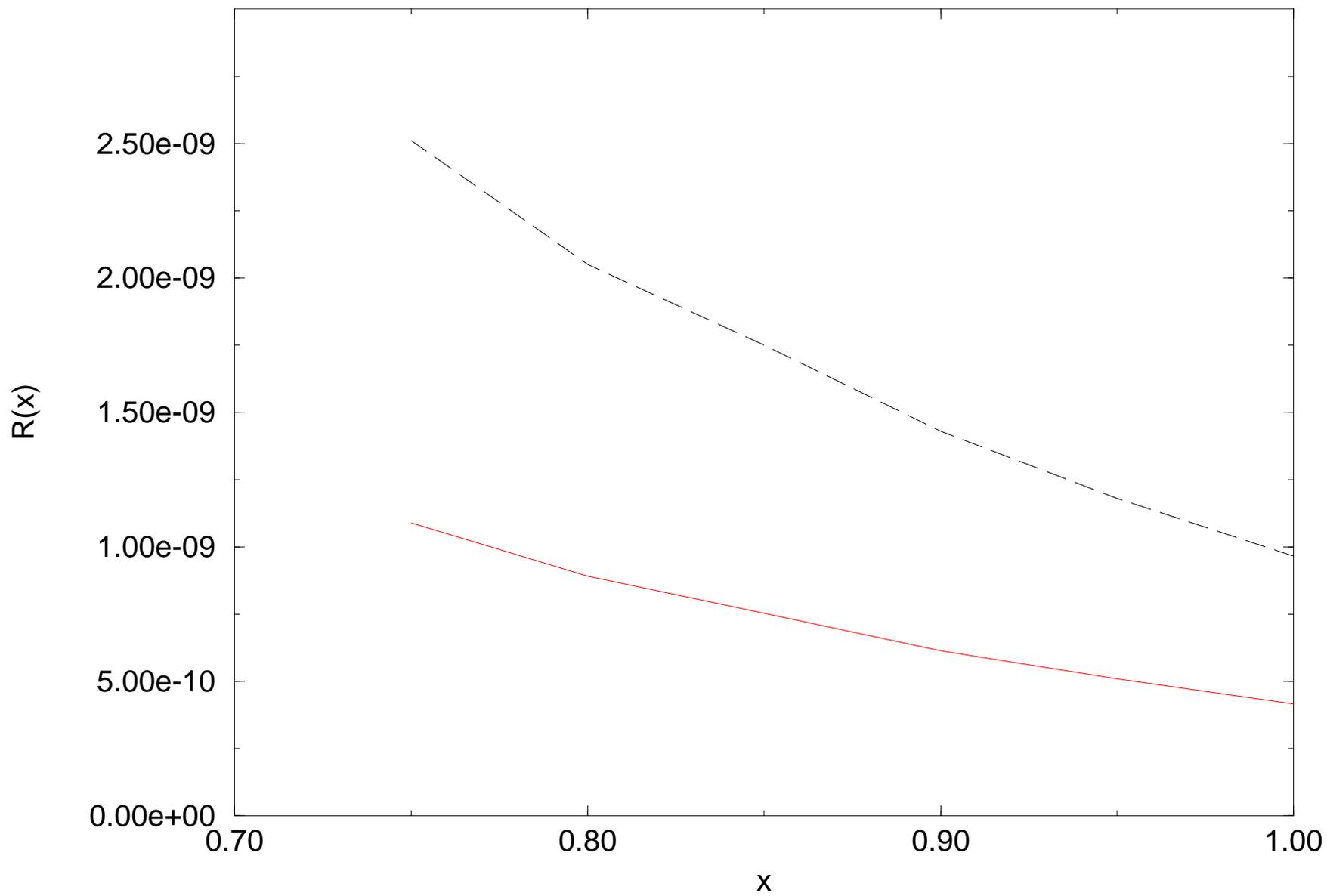
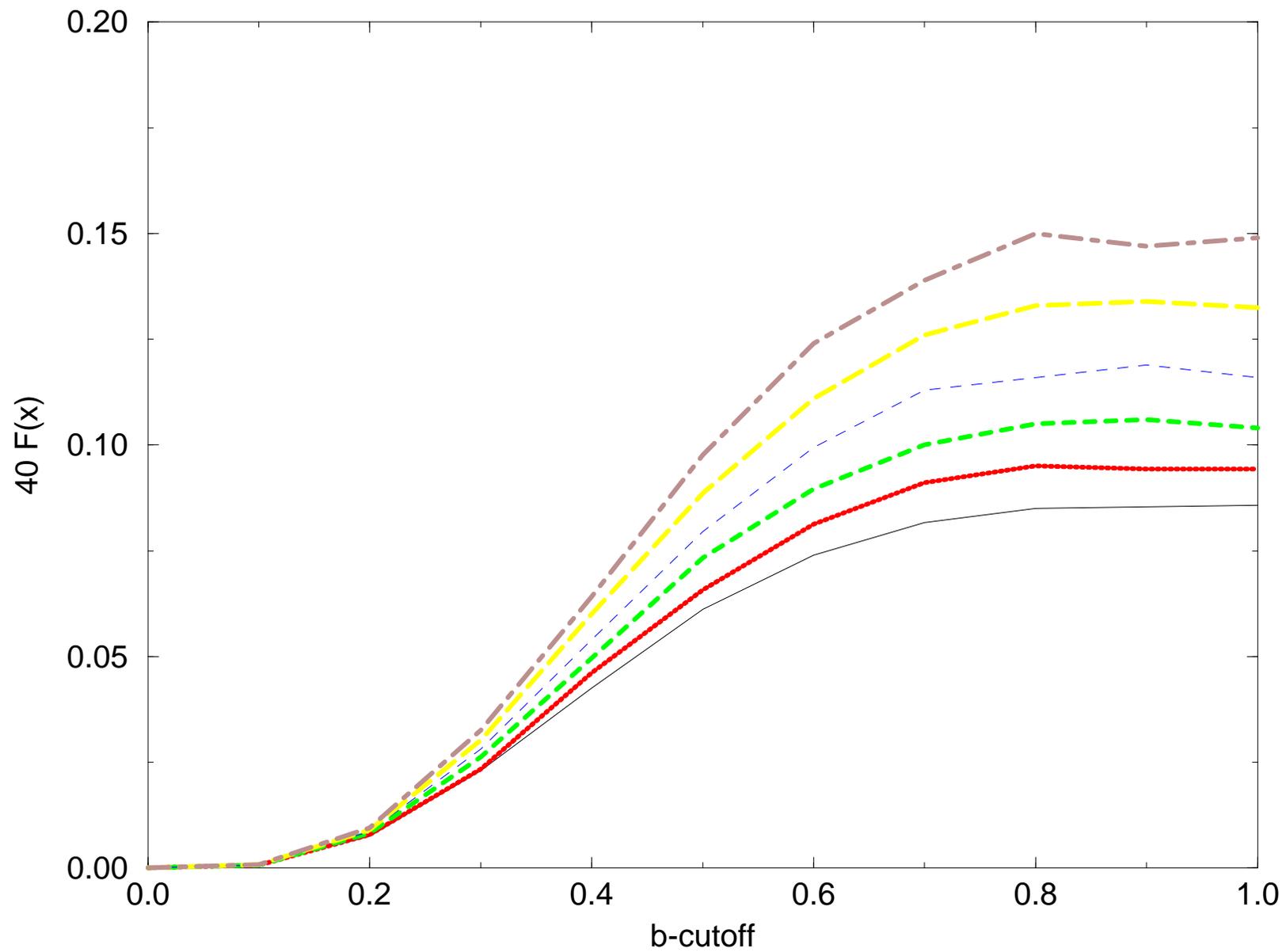
\end{document}